\documentclass[aip,preprint,showpacs,superscriptaddress]{revtex4-1}
\usepackage{graphicx,color}
\usepackage{amssymb,amsmath}

\begin{document}

\title{On the electrodynamic model of ultra-relativistic laser-plasma interactions caused by radiation reaction effects}

\author{A.V~Bashinov}
\author{A.V.~Kim}
\affiliation{Institute of Applied Physics, Russian Academy of Sciences, 603950 Nizhny Novgorod, Russia}

\date{\today}

\begin{abstract}{A simple electrodynamic model is developed to define plasma-field structures in self-consistent  ultra-relativistic laser-plasma interactions  when the radiation reaction effects come into play. An exact analysis of a circularly polarized laser interacting with plasmas  is presented. We define fundamental notations such as nonlinear dielectric permittivity, ponderomotive and dissipative forces acting in a plasma. Plasma-field structures arising during the ultra-relativisitc interactions are also calculated. Based on these solutions we show that about 50\% of laser energy can be converted into gamma-rays in the optimal conditions of laser-foil interaction.}
\end{abstract}


\maketitle

\section{Introduction}

Recent dramatic progress in laser technology has given rise to new projects (e.g., ELI, XFEL) which may bring intensities of the order of 10$^{23}$-10$^{24}$ Wcm$^{-2}$ within reach \cite{ring,eli}. This opens up a wide range of possibilities for exploring novel regimes of ultraintense laser-matter interactions \cite{mourou}. The main feature of such interactions is that the  radiation reaction (RR) effects \cite{rohlich,landau,piazza,piazza1,sokolov,art} coming into play may strongly modify  the conventional relativistic laser-plasma interactions. First attempts to consider such ultra-relativistic interactions accounting for the radiation reaction force were recently made by using particle-in-cell (PIC) simulations, particularly showing the qualitative difference between linear and circular polarization of the driving field \cite{zhidkov,naumova,tamburini,kirk,bulanov2}. 

In this paper we focus on fundamentally new features of self-consistent ultra-relativistic laser plasma interactions. We pay particular attention to the circularly polarized lasers when electrodynamic description of the interaction can be simplified essentially, providing exact solutions for plasma-field structures. For those, we derive fundamental definitions, such as nonlinear dielectric permittivity, ponderomotive and dissipative forces acting on an electron at such ultrahigh intensities. Besides, based on these treatments an effective plasma converter of laser energy into gamma rays is also proposed. 

\section{Basic equations}

To provide a self-consistent approach to ultra-relativistic laser plasma interaction by using macroscopic electrodynamic description (see, for example, \cite{jacson}) we should first clearly define the relationship between macroscopic fields, averaged over small physical volume, and local fields, acting on a particle situated at this point. In conventional electrodynamics of a plasma where the radiation reaction effect is not taken into account these forces are equal (see, e.g., \cite{kadomtsev}), and what is also important is that these fields at the given point are produced by all other particles except the one situated at this point. Using this statement we may incorporate the radiation reaction effect into macroscopic electrodynamics just by replacing in the single particle motion
\begin{equation}
\frac{d{\bf p}}{dt}=e{\bf E^{ext}}+\frac{e}{c}{\bf v}\times{\bf H^{ext}}+{\bf f}_R,
\end{equation}
external fields ${\bf E^{ext}}$, ${\bf H^{ext}}$ by macroscopic fields ${\bf E}$, ${\bf H}$, which are produced by all other particles and these fields (as well including external fields) are governed by the Maxwell equations with the corresponding charges and currents; usually the same procedure is used in PIC modeling. Radiation reaction force can be taken in the Lorentz-Abraham-Dirac (LAD) form or Landau-Lifshitz one; this was much discussed in the literature (see, e.g., \cite{rohlich,ilderton}). Detailed comparative analysis of single particle motion in a laser field was done in Ref.~\cite{bulanov1}.
    
For considering the ultra-relativistic laser plasma interaction when the RR effects are of importance we employ the most simple approach based on relativistic hydrodynamic equations describing the motion of a cold electron fluid accounting for the RR force in the Lorentz-Abraham-Dirac (LAD) form:
\begin{gather}
\frac{d{\bf p}}{dt}=\frac{e}{c}\frac{\partial{\bf A}}{\partial t}+e\nabla\varphi-\frac{e}{c}{\bf v}\times\nabla\times{\bf A}+{\bf f}_R, \label{eq1}\\
{\bf f}_R=\frac{2e^2\gamma^2}{3c^3}\left\{\frac{d^2{\bf v}}{dt^2}+\frac{3\gamma^2}{c^2}({\bf v}\frac{d{\bf v}}{dt})
\frac{d{\bf v}}{dt}+\frac{\gamma^2{\bf v}}{c^2}\left[{\bf v}\frac{d^2{\bf v}}{dt^2}+\frac{3\gamma^2}{c^2}({\bf v}\frac{d{\bf v}}{dt})^2\right]\right\}, \label{LAD}\\
\frac{d N_e}{d t}+ N_e\nabla\cdot{\bf v}=0\label{cont}, 
\end{gather}
where,  instead of electric and magnetic fields, we introduced vector ${\bf A}$ and  $\varphi$ scalar potentials, ${\bf E}=-\frac{1}{c}\frac{\partial{\bf A}}{\partial t}-\nabla\varphi$, ${\bf H}=\nabla\times{\bf A}$, $d/dt=\partial/\partial t+({\bf v}\nabla)$ is the full time derivative, $-e$, $m$ are the electron charge and mass, respectively, ${\bf p}=\gamma m{\bf v}$ is the electron momentum, ${\bf v}$ is the electron velocity, $\gamma=(1-{\bf v}^2/c^2)^{-1/2}$ is the relativistic factor, $c$ is the speed of light, $N_e$ is the electron density. This approach is also justified by the fact that, at extremely high intensities of interest RR acts as a cooling mechanism (see, for example, \cite{tamburini,berezh, mahajan2}) essentially lowering electron temperature as compared to the case when RR is not included. Adopting the Coulomb gauge ($\nabla\cdot{\bf A}=0$), ${\bf A}$ describes the vortex electromagnetic field, and $\varphi$ describes the electrostatic field due to charge separation in the plasma for which the Maxwell equations reduce to 
\begin{gather}\label{wave}
\nabla^2{\bf A}-\frac{1}{c^2}\frac{\partial^2 \bf A}{\partial t^2}=\frac{1}{c}\frac{\partial\nabla \varphi}{\partial t}+\frac{4\pi}{c}eN_e{\bf v},\\
\nabla^2\varphi=4\pi e(N_e-N_0). \label{puisson}
\end{gather}
Here $N_0$ is the unperturbed plasma density. The fluid model, representing a significant simplification over a full kinetic treatment based on particle-in-cell simulations or the Vlasov equation but retaining enough physics to be qualitatively and quantitatively useful\cite{mahajan,berezh}, without the RR force was widely used in the conventional relativistic interactions, particularly to study fundamentally new properties of the interaction. The three well known examples related to the case of circularly polarized electromagnetic wave are: (i) the propagation of intense plane wave due to the relativistic mass correction \cite{akhiezer,sarachik},  (ii) exact solutions for plasma-field structures in the regime of relativistic self-induced transparency (SIT) describing the penetration of incident radiation into overdense plasmas \cite{marburger,tushentsov}, and (iii) electromagnetic solitons of relativistically strong laser fields propagating in underdense and also in overdense plasmas \cite{litvak,kaw}.  The advantage of using circular polarization is that the ponderomotive force has no oscillating (with double laser frequency) component and therefore pushes electrons steadily without significant electron heating. Recently by taking this advantage several schemes for laser ion acceleration into GeV energy range were proposed \cite{rpa,kor}.

\begin{figure}[htpb]
\centering
\includegraphics[width=0.35\textwidth,angle=0]{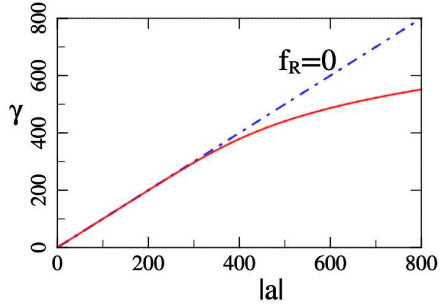}
\caption{(color online) Relativistic factor as a function of field amplitude with (red) and without (blue) RR force.} \label{gamma}
\end{figure}

Next we will also use this advantage but consider the case of importance of RR effects. Paying particular attention to the one-dimensional problem and quasisteady model as well, we assume that the longitudinal motion is frozen, so that in a circularly polarized field ${\bf A}=Re[A(z)({\bf e}_x+i{\bf e}_y)exp(i\omega t)]$ ($\omega$ is the laser frequency) the electron velocity can also be taken in the form ${\bf v}=Re[v(z)({\bf e}_x+i{\bf e}_y)exp(i\omega t)]$ \cite{zeldovich}, where $A(z)$, $v(z)$ are the complex functions. Substituting them into Eq.~(\ref{LAD}) for radiation reaction force we obtain 
\begin{equation}
f_R(z)=-\delta\gamma^4m\omega v(z),
\end{equation}
where the parameter responsible for the RR effects is $\delta=2e^2\omega/3mc^3$, which is approximately equal to 10$^{-8}$ for a 1 $\mu$m wavelength. It is worth noting that similar procedure can also be applied by considering the friction force in the form of Landau-Lifshitz \cite{landau} instead of the LAD form as in Eq.~(\ref{LAD}), but for the latter algebraic operations are much simpler. And as well, in spite of the fact that the LAD force may lead to unphysical self-accelerated solutions for the single-electron interaction problem, in our case of a harmonic process no qualitatively new solutions are added. For convenience of notation let us introduce the following dimensionless variables and parameters: longitudinal coordinate (along the propagation direction) $\xi=z\omega/c$, vector $a(\xi)=eA/mc^2$ and scalar $\phi(\xi)=e\varphi/mc^2$ potentials, $n_e(\xi)=N_e/N_0$ and $n_0=4\pi e^2N_0/m\omega^2$ is the ratio of unperturbed plasma density $N_0$ over the critical one at given laser frequency, the ratio $v/c$ will be denoted as $v$. In this case Eq.~(\ref{eq1}) on the transverse projection gives
\begin{equation}
a(z)=\gamma(1-i\delta\gamma^3)v(z)
\label{a1}
\end{equation}
whereas on the longitudinal one it arrive at  
\begin{equation}
\phi'=\frac{1}{2\gamma a}\left[\frac{(\mid a\mid^2)'}{1+\delta^2\gamma^6 }+i\delta\gamma^3a\frac{aa^{*}{'}-a^{*} a'}{1+\delta^2\gamma^6} \right],\label{force}
\end{equation}
where the prime denotes the derivative with respect to $\xi$. It should be emphasized that equation (\ref{force}) indicates that in the region where the electron density $n(\xi)\ne0$  the force of a longitudinal field due to charge separation is compensated by a sum of two forces; the first one is the ponderomotive force \cite{litvak} and the second one is a dissipative force due to the RR effect, which is generally caused by the imaginary part of dielectric permittivity. The medium absorbs laser photons thus additionally acquiring an impulse along laser propagation direction. From Eq.~(\ref{a1}) we can also obtain an explicit expression of relativistic factor $\gamma$ as a function of field amplitude $\mid a\mid $. Multiplying this equation by the complex conjugate term we arrive at the fourth order algebraic equation with respect to $\gamma^2$ (see also \cite{bulanov1, bulanov}):
\begin{equation}
\delta^2\gamma^8-\delta^2\gamma^6+\gamma^2-(1+|a|^2)=0,
\end{equation}
where we used $|v|^2=\left(\gamma^2-1\right)/\gamma^2$.  This equation has two complex conjugate roots, one is negative and the positive root, suitable for $\gamma^2$ only, is given below:
\begin{equation}
\gamma^2=\frac{1}{4}-\frac{1}{2}B^{1/2}+\frac{1}{4}\left(3-4B+\frac{8-\delta^2}{\delta^2B^{1/2}}\right)^{1/2}  \label{gama}
\end{equation}
where
\begin{gather*}
B=\frac{1}{4}-\frac{q}{\delta^2}[s+(s^2+54q^3)^{1/2}]^{-1/3}+\frac{[s+(s^2+54q^3)^{1/2}]^{1/3}}{ 2^{1/3}3\delta^2}, \\
q=2^{1/3}\delta^2(3+4\mid a\mid ^2), 
s=27\delta^2[1-\delta^2(1+\mid a\mid ^2)]. \nonumber 
\end{gather*}
For clarity, this function is depicted in Fig.~\ref{gamma} for both cases with and without the RR force clearly indicating lowering of the energy of the electron due to its radiation losses at $a>350$. 

Similar way we rewrite Eqs.~(\ref{wave}) and (\ref{puisson}) arriving at 
\begin{gather}
a''+\left[1-\frac{n_0n_e}{\gamma (1-i\delta\gamma^3)}\right]a=0,\label{a}\\
\phi''=n_0(n_e-1).\label{phi}
\end{gather}
Thus the set of equations (\ref{force}), (\ref{a}), (\ref{phi}) together with Eq.~(\ref{gama}) represent the full set of self-consistent ordinary equations describing plasma-field structures that can be realised during ultra-relativistic laser plasma interaction. It should be noted that ion dynamics is neglected here. Although the main reason is simplicity, at extreme field intensities in plasmas as high as 10$^{23}$ Wcm$^{-2}$ electron dynamics, as follows from simulations, is much faster than ion dynamics.

Summarizing the above consideration we underline that in comparison with the conventional relativistic interaction ($f_R=0$), the set of equations  (\ref{force}), (\ref{gama}) - (\ref{phi}) has two fundamentally new features. The first one follows from Eq.~(\ref{force}) describing the laser-plasma equilibrium when the quasistatic force due to charge separation (left side) compensates the action from the laser wave comprising the ponderomotive force (the first term in the right-hand side) and a new one, dissipative force arising from the laser energy absorption in the RR processes (the second term). In the case of single electron motion the latter force is responsible for electron acceleration due to the RR effect (see, e.g., \cite{fradkin}). It should be noted that the right-hand side of Eq.~(\ref{force}) is an exact expression for ponderomotive and dissipative forces arising from the laser fields acting onto the plasma. The second feature is in the wave equation (\ref{a}) in which there appeared an imaginary part, which indicates the importance of absorption processes due to the RR effects.

Let us now apply these equations to particular problems of interest.  

\section{Homogeneous plasma}

To begin with, we consider a very instructive example of a homogeneous plasma ($n_e=1$). In this case, considering equation (\ref{a}) for the laser field only and assuming $a\sim ae^{i\kappa\xi}$ we can obtain the standard nonlinear dispersion relation $\kappa^2=\omega^2\varepsilon/c^2$ (where $\kappa$ is the wavevector of a plane electromagnetic wave) with the following expression for dielectric permittivity
\begin{equation}\label{epsilon}
\varepsilon=1-\frac{n_0}{\gamma (1+\delta^2\gamma^6)}(1+i\delta\gamma^3),
\end{equation}   
which contains real and imaginary parts. At $\delta \to 0$, as is expected, the imaginary part goes to zero whereas the real one reduces to the conventional relativistic dielectric permittivity $\varepsilon=1-n_0/(1+\mid a\mid ^2)$ \cite{akhiezer}. The qualitative contribution of the RR effects is in the imaginary part, i.e., in the plasma conductivity $\sigma=\frac{\omega}{4\pi}Im\varepsilon=\frac{\omega}{4\pi}\frac{n_0\delta\gamma^2}{(1+\delta^2\gamma^6)}$ that appeared due to laser energy absorption caused by the re-emission of photons. For estimates, the imaginary part is comparable with the plasma contribution to the real part ($Im\varepsilon\sim Re\varepsilon-1$) at $\gamma\sim\delta^{-1/3}$, i.e., at $a\sim700$. 

By introducing a single particle scattering cross-section through its definition as the energy absorbed in a volume unit, $\sigma\mid{\bf E}\mid^2$, being equal to the incident flux energy, $c\mid{\bf E}\mid^2/4\pi$,  multiplying by $\sigma_RN_e$, where $N_e$ is the number of scatters, i.e., electrons, we can easily obtain an exact expression for the relativistic Tompson scattering cross-section $\sigma_R$:
\begin{equation}
\sigma_R=\sigma_T\frac{\gamma^2}{1+\delta^2\gamma^6},
\end{equation}
where $\sigma_T=8\pi r_e^2/3$ ($r_e=e^2/mc^2)$ is the electron radius) is the Tompson scattering cross-section. If in the $\mid a \mid<<\delta^{-1/3}$ range we obtain the well-known formula $\sigma_R=\sigma_T(1+\mid a\mid^2)$, whereas for a much higher amplitude $\mid a \mid>>\delta^{-1/3}$ the scattering cross-section decreases as $\sigma_R\approx\sigma_T/(\delta\mid a\mid)$. 

\section{Plasma-field structures in nonhomogeneous plasmas}

\subsection{Semi-infinite plasma}

\begin{figure}
\centering
\includegraphics[width=0.35\textwidth,angle=0]{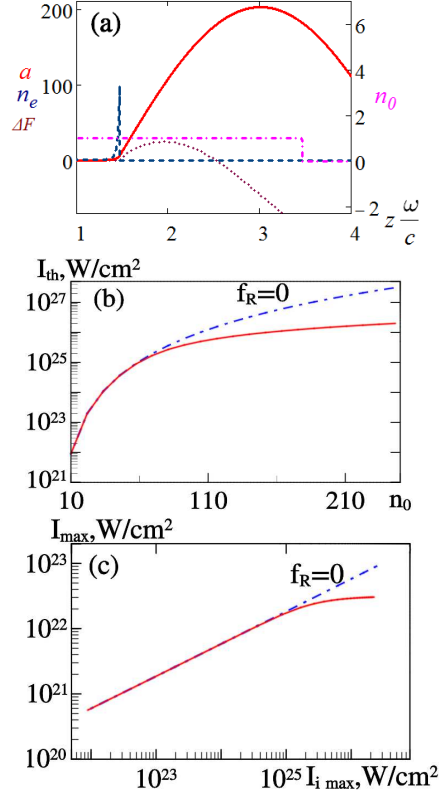}
\caption{(color online) (a) Plasma-field structures for the case of electromagnetic wave incident from the right onto overdense semi-infinite plasma: red - field distribution, blue and violet - electron and ion density distributions, brown dotted line - the difference of the forces acting on probe electron from the laser and quasistatic field due to charge separation. (b), (c) The threshold of induced transparency and the maximum of the laser field attainable in the plasma with (blue) and without (black) RR force, respectively} \label{semi}
\end{figure}

The next also very instructive example is the interaction of electromagnetic waves with overdense plasmas, which are of interest from fundamental point of view and for applications. The most striking effect is the so-called relativistic self-induced transparency (RSIT) effect when an incident relativistically strong electromagnetic wave is able to penetrate deep into overdense plasmas, which are otherwise opaque for low intensity waves \cite{akhiezer,marburger}. To study this effect at ultra-relativistic intensities at which the RR effects should be taken into account we will use Eqs.~ (\ref{force}), (\ref{gama}) - (\ref{phi}) where we assume $a=ue^{i\theta}$. Then equation (\ref{a}) for amplitude splits into two equations for the real functions $u(\xi)$ and $\theta(\xi)$
\begin{gather}    
u''-(\theta')^2 u+\left[1-\frac{n_0n_e}{\gamma(1+\delta^2\gamma^6)}\right]u=0, \label{u}\\
(u^2\theta')'=\delta\gamma\frac{n_0n_e u^2}{1+\delta^2 \gamma^6}, \label{flux}
\end{gather}
which constitute together with Eqs.~(\ref{force}), (\ref{gama}) and (\ref{phi}) a self-consistent set of ordinary equations. It is useful to note that Eq.~(\ref{flux}), expressing the electromagnetic energy conservation law, describes the energy flux decreasing along the propagation path due to radiation losses. This set of equations can be solved numerically assuming for the case of semi-infinite plasma that at $\xi\to -\infty$ the system is unperturbed, i.e., $u,u',\theta,\theta',\phi,\phi'\to 0$, and $n_e\to 1$ \cite{tushentsov}. An example of the plasma-field structures which describe the plasma electron compression  and formation of nonlinear skin-layer distribution for the laser field is shown in Fig.~\ref{semi}(a). An auxiliary curve (brown dotted line), that is the difference of the forces $\Delta F$ acting on a probe electron from the laser and quasistatic field due to charge separation, help us to understand the physics behind the RSIT effect. In fact, at increasing incident intensities this difference of the forces acting on boundary electrons is positive and all electrons are pushed deeper into plasma. However, the point where $\Delta F$ is equal to zero is moving to the boundary point. In the limiting case when this point exactly coincides with the boundary point corresponds to the threshold of the RSIT effect \cite{eremin}. Indeed, at higher incident intensities exceeding this threshold, skin-layer solutions do not exist any longer. Moreover, boundary electrons are in unstable position, since small displacements from their position make the difference of the forces acting on them negative, which means that they will move towards the incident wave, thus allowing the wave to penetrate deeper into the plasma. In Fig.~\ref{semi}(b) we present the RSIT threshold as a function of plasma density with and without the RR effects that shows a decrease in the threshold when the friction force is taken into account. Physically this occurs due to arising of an additional, dissipative force [second term in the right-hand side of Eq.~(\ref{force})] giving the possibility to support such structures with lower incident intensities.

\subsection{Plasma layer}

\begin{figure}
\centering
\includegraphics[width=0.35\textwidth,angle=0]{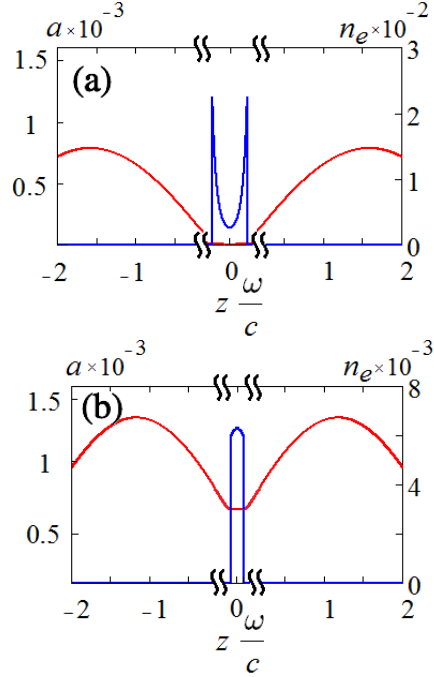}
\caption{(color online) Field (red) and electron density (blue) distributions for 500 nm gold foil symmetrically irradiated by incident intensities of (a) $5\times 10^{23}$ Wcm$^{-2}$ and (b) $3\times 10^{24}$ Wcm$^{-2}$.} \label{foil}
\end{figure}

As the RR effects at ultrahigh laser intensities are directly connected with the generation of high energy photons, the very intriguing possibility is to use them in order to create very efficient converter of laser energy into gamma-rays. However, as shown in simulations \cite{tamburini} and also clearly seen in Fig.~\ref{semi}(c) at incident intensities high enough to generate high energy photons by a single electron, the laser field inside the plasma is much smaller than the incident one. This happens due to the electrodynamic effect of laser interaction with plasma, i.e., electron peaking by laser compression strongly reflects the incident wave resulting in appreciable decrease of electromagnetic fields in the plasma as shown in Fig.~\ref{semi}(c), thereby strongly suppressing the RR effects. However, this obstacle might be overcome by using thin plasma layers (thin foils) and symmetrical irradiation from both sides. First, we should optimize foil parameter so that the laser field should be able to penetrate deep into electron region and, second, symmetrical irradiation should arrest longitudinal electron motion. To do so, we will solve numerically Eqs.~(\ref{force}), (\ref{gama}),  (\ref{phi}),(\ref{u}), and (\ref{flux}) in the class of symmetrical solutions, i.e., assuming that $\xi=0$ $u'=\theta'=\Phi'=0$ \cite{kor1}. The typical plasma-field structure realized during the interaction is shown in Fig.~\ref{foil}(a) that models the symmetrical irradiation of gold foil ($n_0=500$) of 500 nm width. We see that in spite of high incident intensity of $5\times10^{23}$ Wcm$^{-2}$ the field in the electron region is relatively small ($a_{max}\sim 5$). Strong reflection of the incident wave from the compressed electron layer (compression factor is about 75 and the maximum electron density increases by several orders of magnitude reaching $2\times10^{26}$ cm$^{-3}$) greatly suppresses the RR effects, thus gamma-rays are not generated in fact. Whereas by carefully choosing foil parameters, such as plasma density and width with respect to the given laser intensity, plasma-field structures can be favourably modified as depicted in Fig.~\ref{foil}(b). In this case, for example, for an incident intensity of $3\times10^{24}$ Wcm$^{-2}$ and the same foil parameter as in  Fig.~\ref{foil}(a), being a single hump the electron layer experiences strong enough electromagnetic field for the RR effects to come into play. In fact, electron layer compressed by a factor of $2.5\times10^3$, reaching a very narrow width of $\lambda/5000$ and density of $2\times10^{27}$ cm$^{-3}$, is able to convert 45\% of incident energy into high energy photons of 45 MeV.  

\begin{figure}
\centering
\includegraphics[width=0.5\textwidth,angle=0]{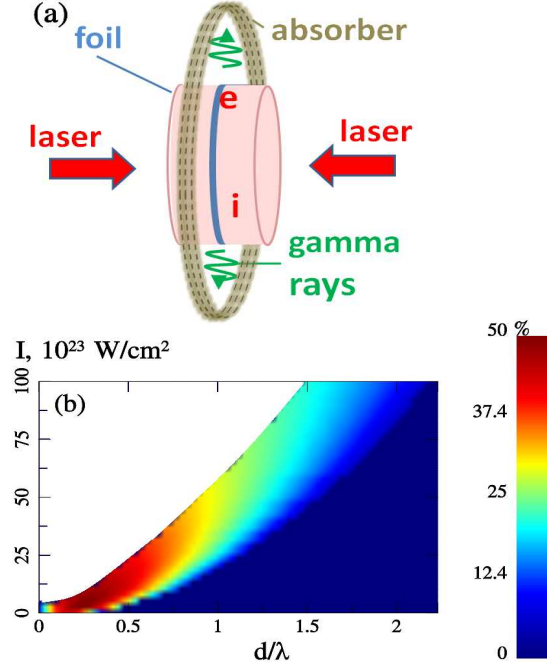}
\caption{(color online) (a) Schematic of the laser-foil converter; orange - ion density, blue - compressed electron density, and (b) the energetic efficiency of the method proposed.} \label{proposal}
\end{figure}

Based on the above results we may propose the following scheme of very efficient laser energy conversion into gamma-rays. A thin foil target, preferably consisting of heavy ions, should be irradiated symmetrically from opposite directions by counter-propagated laser pulses with ultra-relativistic intensities of about $10^{23}$-$10^{24}$ W/cm$^{-2}$ or more. By properly choosing the foil thickness at the given incident intensity, all compressed electrons will synchronically rotate in a narrow plane and produce synchrotron emission along instantaneous velocity. The diagram of radiation, rotating within this plane too, can also be very narrow in proportion to $\gamma^{-1}$. Such a very important issue of the  gamma-ray source, which is essentially two-dimensional, can be used to collect high energy photons for applications by surrounding the gamma-ray radiated narrow electron layer by respective consumers as shown schematically in Fig.~\ref{proposal}(a). To show the efficiency of the method proposed we have performed a parametric scan over a range of foil parameters and laser intensities and calculated the energy absorbed due to RR effects over the electron layer as follows from Eq.~(\ref{a}) as $Q=\frac{\omega}{4\pi}\int\text{Im}\varepsilon\mid a\mid^2d\xi=\frac{\delta\omega n_0}{4\pi}\int\frac{n\gamma^3\mid a\mid ^2}{\gamma(1+\delta^2\gamma^6)}d\xi$, where $\varepsilon$ is defined by Eq.~(\ref{epsilon}). In Fig.~\ref{proposal}(b) we present the conversion efficiency defined as $\eta=\frac{Q}{2I_i}$ ($I_i$ is the incident intensity) as a function of the width of a gold foil that shows the maximum efficiency of 50\%, which means that half of the laser energy can be converted into gamma-rays.       

\section{Conclusions}

In summary, we have studied the fundamental issues of ultra-relativistic laser-plasma interactions when the RR effects can play a crucial role. We paid particular attention to the case of circularly polarized lasers when electromagnetic forces acting along the propagation direction steadily shift electrons forwards, thus making the quasistationary approach suitable for analytical treatment. It allows us to give fundamental definitions such as nonlinear dielectric permittivity, ponderomotive and dissipative forces acting on a plasma in exact form. We have also determined plasma-field structures arising during ultra-relativistic interactions and shown that in optimal conditions of laser-foil interaction about 50\% of laser energy may be converted into gamma-rays.   

This work was supported in part by the Ministry of Education and Science of the Russian Federation under Contract No. 11.G34.31.0011.

\end{document}